# Experimental unsaturated soil mechanics

Pierre Delage
*Ecole Nationale des Ponts et Chaussées, CERMES (Geotechnical Dpt), Paris, France*

ABSTRACT: In this general report, experimental systems and procedures of investigating the hydro-mechanical behaviour of unsaturated soils are presented. The water retention properties of unsaturated soils are commented and linked to various physical parameters and properties of the soils. Techniques of controlling suction are described together with their adaptation in various laboratory testing devices. Some typical features of the mechanical behaviour of unsaturated soils are presented within an elasto-plastic framework.

An attempt to describe the numerous and significant recent advances in the investigation of the behaviour of unsaturated soils, including the contributions to this Conference, is proposed.

## 1 INTRODUCTION

Researches in unsaturated soil mechanics considerably developed in the past decades, through the simultaneous development of experimental investigations and theoretical analyses. It is worth reminding that geotechnical research in unsaturated soils started from significant contributions from researchers in Soil Science, in the field of water retention and water transfer, as shown in the comprehensive historical analysis of Barbour (1998). It is the case, among other things, of the concept of water potential (Buckingham 1907), of the development of the equation for liquid water flow in unsaturated soils (Richards 1928), of the filter paper technique of measuring suction (Gardner 1937) and of the pressure plate apparatus for controlling suction in soils (Richards 1941) that formed the base of the axis translation method (Hilf 1956). Similarly, Gardner (1956) and Corey (1957) respectively proposed a transient and a steady state method of measuring the water permeability coefficient as a function of the degree of saturation.

Geotechnical engineers started at that time to investigate the water retention properties of soil (Croney 1952) and a synthesis of pioneering researches in unsaturated soils was made in 1961 (Conference on Pore pressure and suction in soils, London). A significant amount of information was presented in a paper on the control and measurement of suction in the laboratory by Croney et al. (1952). The techniques presented and used were the suction plate method, the tensiometer, the centrifuge method, the pressure membrane apparatus, the freezing point depression method and the vacuum dessiccator (in which the relative humidity was controlled using saline solutions or sulphuric acid). Measurements of suction using the electrical resistance of porous gauges made up of plaster of Paris or alumina cement ("gypsum block" type), that were previously calibrated as a function of their water retention curve, were also presented. Most of these techniques of control and measurement of the suction are presently commonly used and some of them have been significantly improved, as will be discussed later. Finally, the only technique of controlling suction not presented in Croney et al. (1952) is the osmotic technique, also developed by soil scientists (Zur 1966) and further adapted to geotechnical engineering by Kassiff & Ben Shalom (1971).

As expected, geotechnical engineers were interested by the application of stresses to unsaturated soils. In this regard, the first suction controlled triaxial apparatus, developed by Bishop & Donald (1961) (Figure 1) is of particular historical and scientific interest. This adaptation of the axis translation method to the triaxial apparatus was later on adopted by most researchers carrying out triaxial tests on unsaturated soils. The delicate problem of the control of volume change of the sample, that will be further discussed, was solved by optically monitoring the level of the mercury contained in a glass cylinder surrounding the sample. Water exchanges during shearing were also monitored, and

Bishop and Donald presented the complete set of data that characterise the mechanical behaviour of an unsaturated soil during a constant water content shearing test, i.e. the changes in :

- net mean stress $(p - u_a)$
- deviatoric stress $q$
- suction $u_a - u_w$
- volume $V$
- degree of saturation $S_r = V_w/eV_s$ (where $V_w$, $e$ and $V_s$ are respectively the water volume, void ratio and volume of solid of the sample).

Changes in these parameters during shearing at a constant confining pressure $\sigma_3$ were monitored as a function of the axial strain $\varepsilon_1$.

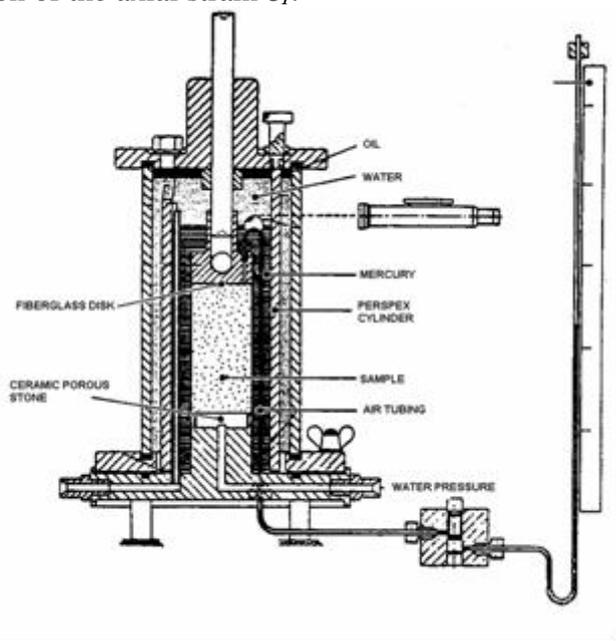

Figure 1 : Suction controlled triaxial apparatus (Bishop & Donald 1961).

The relevant set of independent stress state variables to be used in unsaturated soil mechanics started to be investigated at the same time by Coleman (1962), Matyas & Radhakrisna (1968), Fredlund & Morgenstern (1977) and more recently by Tarantino et al. (2000). In an axisymetric stress of stress, further analysis showed that the work conjugate variables to consider (Edgar 1993, Wheeler & Sivakumar 1995, Houlsby 1997, Dangla et al. 1997, Romero & Vaunat 2000) are the following :

- net mean stress  $\quad$ - volume
  $p - u_a$  $\quad\quad\quad\quad\quad\quad$ $\delta\varepsilon_v = \delta\varepsilon_a + 2\delta\varepsilon_r$
- deviatoric stress $\quad$ - shear strain
  $q$  $\quad\quad\quad\quad\quad\quad\quad$ $\delta\varepsilon_s = 2(\varepsilon_a - \delta\varepsilon_r)/3$
- suction  $\quad\quad\quad\quad\quad$ - water content
  $u_w - u_a$  $\quad\quad\quad\quad\quad$ $\delta\varepsilon_w = -\delta e_w/(1 + e_0)$

Following Bishop & Donald (1961)'s paper, further investigations on the shear strength of unsaturated soils were carried out using the direct shear box, in which the control of suction was made using the axis translation method (Escario 1980, Gan et al. 1988). At that time, more attention was paid to the investigation of volume change behaviour using controlled suction isotropic cells or oedometers (Matyas & Radhakrishna 1968, Barden et al. 1969). Papers presenting data from controlled suction triaxial testing were presented later (Gulhati & Satija 1981, Ho & Fredlund 1982a, Josa et al. 1987, Delage et al. 1987). The use of the controlled suction triaxial apparatuses further developed in the 90's, together with the developments of the first critical state constitutive model (Alonso et al. 1990).

Relatively few data on total volume changes and water exchanges during shearing, indispensable for a full understanding of the constitutive behaviour of unsaturated soils, were available in the literature until recently (Wheeler & Sivakumar 1995). Research efforts were rather aimed at determining maximum shear resistance of unsaturated soils for conventional geotechnical analysis, mainly using the direct shear box (Escario & Saez 1986). It is thought however that the growing use of finite elements codes for helping in the resolution of practical geotechnical problems will necessitate more experimental constitutive data on unsaturated soils under controlled suction. As can been seen in this Conference and in recently published papers that will be mentioned further on, an important investigation effort is presently being carried out in this domain.

The growing interest in unsaturated soils mechanics has been concretised by the initiation of a series of International Conferences specifically devoted to unsaturated soils, started in Paris (Alonso & Delage 1995), prolonged in Beijing (Liu 1998) and being presently held in Recife (Juca et al. 2002). As mentioned by Fredlund (2000), this series of International Conference is included in a context of various Conferences including the series of Expansive Soils Conferences and special sessions or regional Conferences : sessions in the ASCE Geo-Conferences in the USA (Houston & Wray 1993, Houston & Fredlund 1997, Shackelford et al. 2000), Unsaturated soils Symposia in Brazil, International Workshop in Trento (Italy) in 2000 (Tarantino & Mancuso 2000), Unsat-Asia 2000 Conference in Singapore (Rahardjo et al. 2000). Presently, various planned events are the Unsat-Asia 2003 Conference (Kobe, Japan, 2003), an International Workshop on Unsaturated Soils (Napoli, Italy, 2004) and the 4[th] International Conference on Unsaturated Soils in Phoenix (Arizona) in 2006. Special Issues of Journals have also recently been dedicated to unsaturated soils and swelling soils : *Engineering Geology* (1999), *Geotechnical and Geological Engineering* (Toll, 2001) and a Symposium in Print of *Géotechnique* is planned in 2003.

Various keynote lectures and state of the art reports were produced, together with the well-known textbook of Fredlund & Rahardjo (1993).They contain a lot of valuable information on unsaturated soil experimental testing and behaviour. Some of them are listed below :

- Special problems soils (Alonso, Gens & Hight 87)
- The scope of unsaturated soil mechanics (Fredlund 95)
- Experimental techniques (Juca & Frydman 1995)
- Suction measurements (Ridley & Wray 95)
- Mechanical behaviour (Delage & Graham 95)
- Modelling of compacted soils (Gens 95)
- Constitutive behaviour (Wheeler & Karube 95)
- Hydraulic conductivity (Benson & Gribb 97)
- Mass transfer in unsaturated soils (Wray 98)
- Modelling expansive soil behaviour (Alonso 98)
- Soil water characteristics curves (Barbour 98)

Valuable information can also been found in textbooks of Soil Physics such as, for example, Hillel (1980).

The Proceedings and other contributions of workers involved in the above-mentioned Conferences represent the findings of a community of researches mainly concerned by traditional civil engineering aspects. It should be mentioned that extensive research that also concerns some aspects of unsaturated soil behaviour have been conducted in relation with geo-environmental problems, such as waste containment and soil pollution. In particular, special attention has been given to the behaviour of compacted expansive soil used as engineered barriers for the isolation of high activity nuclear wastes. In these studies, quite high suctions (up to several hundreds of MPa) and high stress levels (up to 60 MPa) are considered, involving new experimental developments. Thermal and chemical effects on the behaviour of engineered barriers or clay liners are now considered in both waste disposal and soil pollution problems, the emphasis being rather put on mass transfer phenomena. Related publications are sometimes published in specialised Conferences on Environmental Geotechnics, or in special Issues of Journals (Engineering Geology 41, 1-4, 1996 ; 47, 4, 1997 ; 54, 1-2, 1999 ; 64 N° 1, 2002). Special attention is also devoted on microstructure observations related to macroscopic behaviour, in terms of mass transfer and of mechanical behaviour (Shizuoka International Symposium on Suction, Swelling, Permeability and Structure of Clays, Adachi & Fukue 2001). Various papers dealing with these aspects are presented in this Conference. However, the participation from the geo-environmental community still appears relatively modest.

## 2 THE WATER RETENTION CURVES

### 2.1 *Experimental techniques*

A fundamental property of an unsaturated soil related to its ability to attract water at various water contents and suctions is characterised by the water retention curve (WRC), also called soil water characteristic curve (SWCC). Controlled suctions are applied to a sample not submitted to stress, using generally the axis translation technique and a pressure membrane apparatus (Richards 1941). Water content and volume changes are more often monitored by withdrawing the sample from the cell, once equilibrium in suction is reached (generally after 7 days), and samples are weighted and their dimensions manually measured.

An alternative technique of determining the WRC is the osmotic technique (Zur 1966, Kassiff & Ben Shalom 1971), where the soil sample is placed in contact with a semi-permeable membrane behind which an aqueous solution of large sized polyethyleneglycol (PEG) molecules is circulated. Since water molecules can cross the membrane whereas PEG molecules cannot, an osmotic suction, that increases with the PEG concentration, is applied to the soil through the membrane. Note that, since water transfer occur in the liquid phase, the osmotic technique controls the matric suction of the soil, and not the osmotic suction. The system is easy to adapt for the determination of the WRC by placing the sample in a tube-shaped semi-permeable membrane and by plunging it in a PEG solution stirred by a magnetic stirrer (Cui & Delage 1996).

Williams and Shaykewich (1969) found a good agreement between independent calibrations of performed by various authors by measuring or controlling the relative humidities induced by PEG solutions at various concentrations for total suction included between 0 and to 1.5 MPa. More recently, Delage et al. (1998) extended this calibration up to 10 MPa (Figure 2, the left y axis is in a root square scale). The Figure also shows that the calibration is independent of the molecular weight of the PEG (included between 1 500 and 20 000).

By using a null type osmometer, Waldron & Manbeian (1970) obtained however some difference in the calibration data, that were subsequently confirmed by direct suction measurements carried out by Dineen & Burland (1995). These authors evidenced that a membrane effect, that increases with increased suction, could modify the calibration data by reducing the applied suction (100 kPa near 500 kPa and 200 kPa near 1 MPa as compared to Williams & Shaykewich data). This effect was further analysed by Slatter et al. (2000a).

The advantage of the technique is its simplicity and the easiness to reach high suctions in a safe way. A disadvantage is due to the weakness of the membrane and to its sensitivity to bacteria attacks (solved by putting some drops of penicillin in the solution). Another problem during long term hydration tests is the possible PEG crossing through the membrane. This can be solved by using a purifying system (Delage & Cui 2003) that eliminates the smaller PEG molecules contained in the solution.

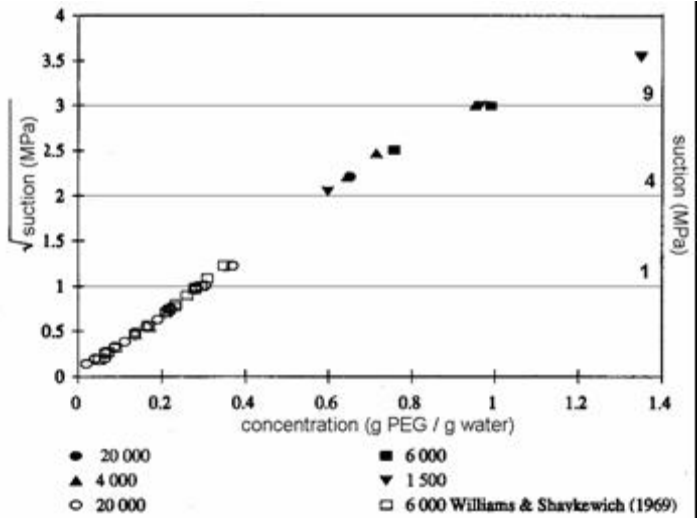

Figure 2 : Extension of the osmotic suction up to 10 MPa (Delage et al. 1998)

A comparison between various existing methods of controlling suction was carried out by Fleureau et al. (1993) on a kaolinite slurry (

Figure 3). A good overall agreement is observed in the drying (saturated) stage, with perhaps some problems in the re-wetting path from large suctions in an unsaturated state, once the air entry value has been reached. This could be related to the problems of the axis translation method in the zone of occluded air (Bocking & Fredlund 1980, Fredlund & Rahardjo 1993). Further investigation on this point is obviously needed.

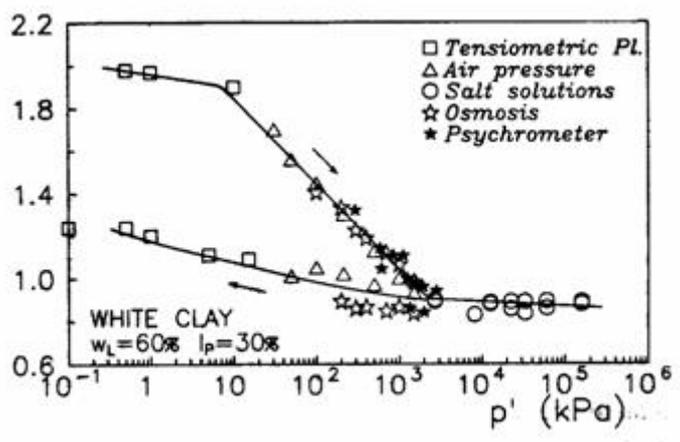

Figure 3 : Comparison between various suction controlled techniques (Fleureau et al. 1993)

Further developments concerning the determination of WRCs concern the automatic monitoring of changes in weight achieved by placing the cell on a precision balance (Takeshita & Kohno 1995).

In this Conference, various WRC determination were presented : *Mata, Romero & Ledesma* carried out WRC determination on bentonite-sand mixture samples by measuring suction with a transistor psychrometer between 0.5 and 70 MPa. *Melgarejo, Ridley & Dineen* used filter paper measurements whereas *Côté, Konrad & Roy* used tensiometer suction measurements together with TDR water contents measurements in a 305 mm diameter column ; *Nishimura & Fredlund* controlled high suctions through the relative humidity and *Paronuzzi, Del Fabbro & Maddaleni* used the pressure plate apparatus and tensiometer measurements.

2.2 *Typical water retention curves*

As an example, Figure 4 (Croney 1952) presents the WRC and the shrinkage curve of a plastic clay (55%<2μm, $I_p = 52$, $w_L = 78$) obtained on a large range of suction (controlled with pressure plate apparatus and controlled relative humidity). As compared to less plastic soils, plastic clays swell and shrink when submitted to suction cycles, and the curves does not correspond to a constant condition of density. The first cycle of suction applied on the intact sample brought to a wet state up to 79 kPa shows a closed loop, probably due to the fact that the higher suction experienced by the sample during its geological history was higher than 79 kPa. Conversely, when the intact sample is brought up to $10^3$ MPa (pF 7 - oven drying), the loop is no more closed, showing irreversible microstructure deformation due to the application of suctions that had never been experienced by the soil. Finally, a subsequent suction cycle up to $10^3$ MPa provides a closed loop. The drying sequence of a slurry made up with the same soil shows a similar behaviour above 1 MPa, showing that microstructure effects are erased by the application of high suctions. Observation of the shrinkage curve (volume versus water content) shows that the air entry value is observed at w = 17.5%, at a suction of approximately 2 500kPa, showing that the soil remains saturated up to very high suctions. This point is not apparent when the WRC is plotted in a suction versus water content diagram as in Figure 4, whereas it appears clearly in a diagram giving the suction change versus the degree of saturation (Figure 5).

More recently, a system allowing for WRC determination at various temperatures has been proposed by Romero et al. (2001). To do so,

particular attention was paid to problems related to possible effects of water evaporation.

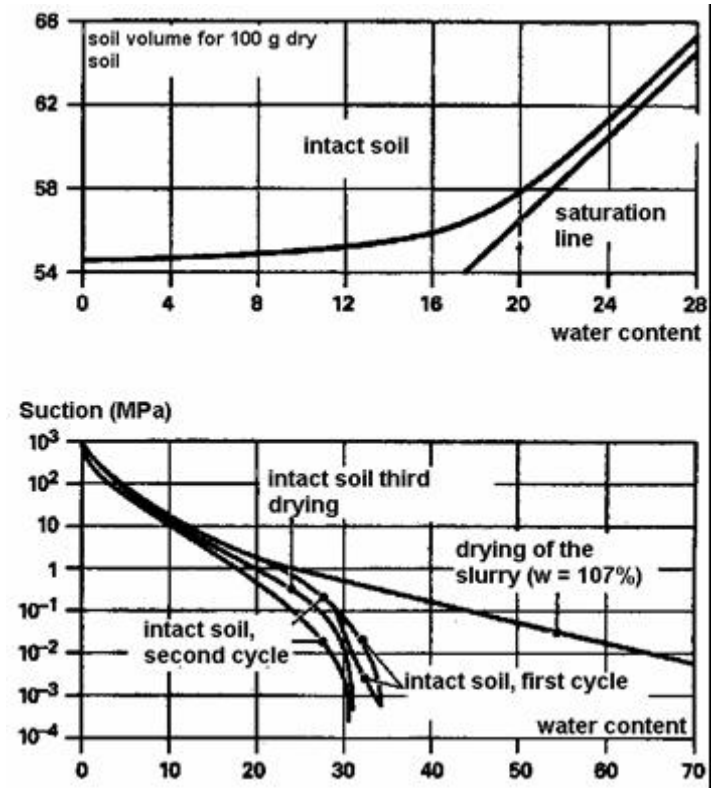

Figure 4 : Shrinkage (a) and water retention curves (b) of a plastic clay (Croney 1952)

WRCs were related to basic soil parameters by many authors (including Hillel 1980, Fredlund & Rahardjo 1993, Fredlund & Xing 1994, and various references presented in Agus et al. 2001 and commented further on). For granular soils, obviously, the WRC can be related to the grain size distribution of the soil. At the same suction, a poorly graded clean sand will retain more water than a well graded sand, due to the existence of smaller pores. Similarly, an increased amount of water will be retained at the same suction in soils containing a larger clay fraction, i.e. that are more plastic (Black 1962 in Delage & Graham 1995, Kassiff et al. 1969) and in a denser soil. In fine-grained soils, the ability of the clay fraction to attract water molecules due to physico-chemical actions in relation with the specific surface and the cations exchange capacity (Mitchell 1993) is quantified by the liquid limit and the plasticity index.

These properties are illustrated in Figure 5 (Barbour 1998, see also Vanapalli et al. 1999) that shows, in another possible presentation of the WRC where the suction axis is taken horizontal, the influence of the nature of the soil, of consolidation (on an initially saturated sample) and of the compaction characteristics. Figure 5a shows that at the same suction (1 MPa for example), the sand is almost dry ($S_r = 10\%$) whereas the degree of saturation of fine grained soils increases with the plasticity ($S_r = 15\%$, 67 and 97% for the Botkin silt, Indian head till and Regina clay respectively).

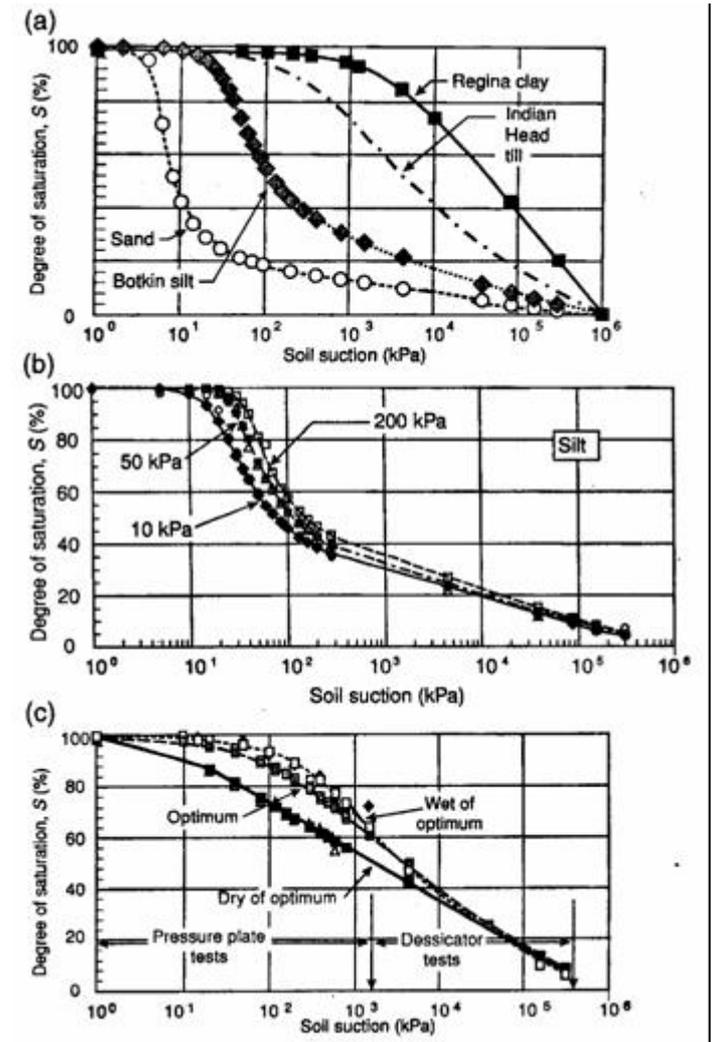

Figure 5 : Influence of a) soil texture, b) consolidation and c) compaction on the water retention properties of soils (from Barbour (1998) and Vanapalli et al. (1999).

2.3 *Relationship between water retention curves and soil characteristics*

2.3.1 *Microstructure*

The effect of consolidation and compaction observed in Figure 5 can be directly related to observations using the scanning electron microscope (SEM) and to changes in the pore size distribution (PSD) curves determined using mercury intrusion. The progressive collapse of the larger existing pores in a saturated soil submitted to increased compression stress (Delage & Lefebvre 1984, Griffiths & Josi 1989, Lapierre et al. 1990) explains the increase of the air entry value (also called bubbling pressure) observed in Figure 5b. Mercury intrusion PSD investigations carried out on compacted soils (Ahmed et al. 1974, Delage et al. 1996) also showed that smaller entrance pore values were observed on soil compacted wet of Proctor optimum. The difference in microstructure between silt samples compacted at the same unit weight (16 kN/m3) dry and wet of optimum are illustrated by the photos in Figure 6 (Delage et al. 1996). The sample compacted dry of optimum is characterised by an aggregate microstructure with large inter-aggregates pores, much larger than that of the matrix microstructure of the "wet" sample, where porosity is defined inside the clay matrix, with significantly smaller intra-matrix pores. This difference explains the difference in air entry value observed in Figure 5c.

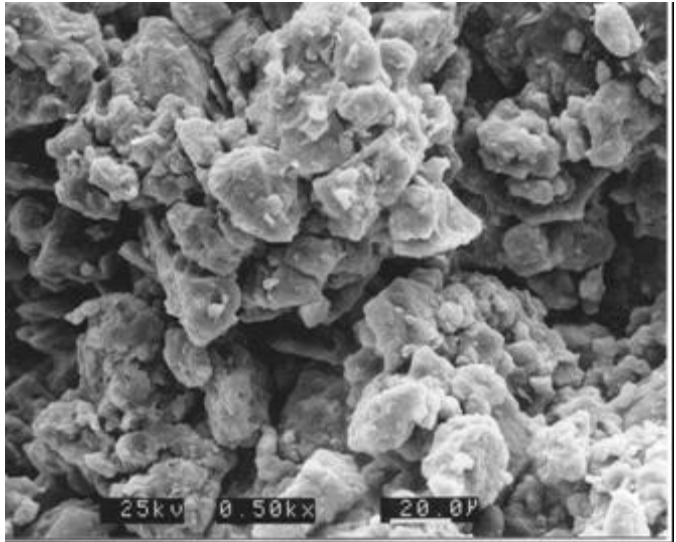
a)

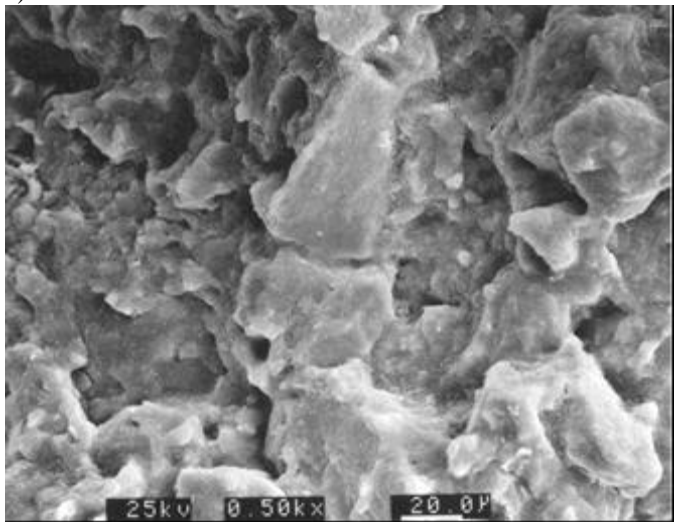
b)
Figure 6 : SEM photos of a compacted Jossigny silt : a) dry of optimum – b) wet of optimum (Delage et al. 1996)

It should be noted however that the bimodal PSD curves that characterise the aggregate microstructure of a soil compacted dry of optimum (Ahmed et al. 1974, Delage et al. 1996) do not correspond to bimodal WRCs. Starting from a saturated state, experience shows that one cannot observe on the WRC any distinction between the draining of the inter-aggregate pores and that of intra-aggregate pores.

Previous comparisons between the WRC and PSD curves of two compacted fine grained soils were presented by Prapaharan et al. (1985). The mercury intrusion curve (where a non wetting fluid intrudes the porous volume) is compared to the main drying curve of the WRC (where air is the non wetting fluid). The comparison was made by considering the ratio between the soil-fluid capillary interactions between mercury (contact angle $\theta = 140°$ and interfacial tension $\sigma_{Hg} = 0.485$ N/m$^{-1}$) and water ($\theta = 180°$ and $\sigma_w = 0.07275$ N/m). In this approach, the mercury intrusion PSD curve can be considered as a retention curve, or the WRC as an air intrusion PSD curve.

A good agreement was observed by Prapaharan et al. (1985) at lower suction. At higher suction, smaller pores were identified by the WRC, as compared to the mercury intrusion PSD curve. Delage et al. (1995) drew similar conclusions on a compacted silt, whereas they observed that a good agreement on all the suction range in an inactive sandstone (inactive in term of mineral-water interaction). Similar observations were made by Romero et al. (1999) and Aung et al. (2001). The main difference between WRC and PSD curves is due to the fact that mercury intrusion occurs in a fixed porous medium, defined in a compacted soil by its density and water content. On the contrary, the WRC is the reflect of the response of a porous medium that changes when the water content changes. SEM observations also showed (Cui 1993, Delage et al. 1996) that the clay fraction in soils compacted at higher moisture contents were much more apparent, due to the significant larger volume of hydrated clay particles. In this regard, the observation of a smaller porosity deduced from WRC analysis is coherent.

Further insight in the relation between WRC, PSD curves and permeability was provided by Romero et al, (1999) on a compacted Boom clay ($w_P = 29\%$, $w_L = 56\%$, 50%<2μm). The compaction diagram presented in Figure 7 shows the density curves obtained at various static compaction stresses, together with the contours of equal suctions. It recalls an important property of compacted soils, already shown by Gens et al. (1995) and Li (1995), illustrated by the vertical lines of equal suction. They confirm that, at lower water contents (<15% for the compacted Boom clay) and for saturation degrees lower than 70%, compression occurs due to the collapse of the macropores of the inter-aggregates porosity. In this domain, the suction is controlled by the water fixed to the clay particles inside the aggregates.

Figure 7: Static compaction curves of Boom clay with contours of equal suction (Romero et al. 1999).

In other words, in this zone, *constant water content tests are also constant suction tests*. This consideration, together with SEM observations and mercury intrusion PSD measurements allowed a more precise definition of the effect of microstructure on the WRC curves of Boom clay compacted at two dry unit weights (w = 15%, $\gamma_d$ = 13.7 and 16.7 kN/m$^3$, also represented in Figure 7 as blank crosses), as shown in Figure 8.

Figure 8: Effect of microstructure on the water retention curves of a compacted Boom clay (Romero et al. 1999)

Based on the determination of the inter-aggregate porosity from PSD curves, the authors demonstrated that the section of the WRC at suctions lower than 2 MPa, where both curves differ, concerned free inter-aggregate water whereas higher suction where both curve superimpose concerned intra-aggregates water. As said before, compaction from the looser sample to the denser one will only concern the inter-aggregate porosity, as was confirmed by the fact that the intra-aggregate porosity as identified by MIP is similar.

2.3.2 *Grain size, plasticity, density*

As commented before, the effect of plasticity on the shape of the WRC was investigated by Black (1962, see also in Delage & Graham 1995) on various British soils. Agus et al. (2001) present a complete study carried out on 8 Singapore soils in which they attempted to relate the WRC to basic parameters of the soils through a multivariate statistical analysis. They refer to previous work of Arya & Paris (1981), Ahuja et al. (1985), Saxton et al. (1986, who worked on a data base of 2541 soils), Fredlund et al. (1997) and Arya et al. (1999). They adopt Fredlund & Xing (1994)'s equation (defined by three constants a, m and n) that they relate to 9 basic parameters: 5 parameters taken from the grain size distribution curve, together with $w_P$, $w_L$, Skempton's activity (*A*) and the dry density. A good fit is statistically obtained on Singapore soils when accounting for the 9 parameters. The results obtained using only 4 parameters are also satisfactory. This approach should now be applied to soil originating from other areas.

In this Conference, various papers presented in the session "*Storage characteristics*" deal with the relationship between soil properties and the WRC. *Mata, Romero & Ledesma* considered chemical effects to further investigate the osmotic component of suction. *Côté, Konrad & Roy* considered the effect of fines contained in different pavement base course granular materials. *Leong & Rahardjo* studied the effect of the energy compaction on the WRC of a residual soil from Singapore. Various papers deal with the retention properties of natural soils. *Vertamatti & Araùjo, Camapum de Carvalho, Cabral Guimarães & Feitonsa Pereira* studied Brazilian soils. *Paronuzzi, Del Fabbro & Maddaleni* investigated the effects of pedogenetic features in soils from unstable colluvial slopes in Italy. *Melgarejo, Ridley & Dineen* used an adaptation of the filter paper technique allowing for the determination of the WRC of a sandy silty clay from Rio.

*Bicalho, Znidarcic & Ho* extended the determined of the WRC curves in the range of negative suction, i.e. positive water pressure (Figure 9), where air continuity vanishes, leading to the occurrence of occluded air bubbles. They obtained a reasonable prediction using a combination of Boyle's law and Henry'law.

A detailed investigation on the status of water in bentonite samples was carried out by *Guillot,*

*Fleureau, Al Mukhtar & Bergaya* using a thermogravimetric analysis (in which the water content loss due to a progressive temperature increased up to 1000°C is carefully monitored) in order to distinguish between free water and adsorbed water. They concluded that free water was expelled below 150°C, whereas temperatures higher than 150°C were necessary to release adsorbed water.

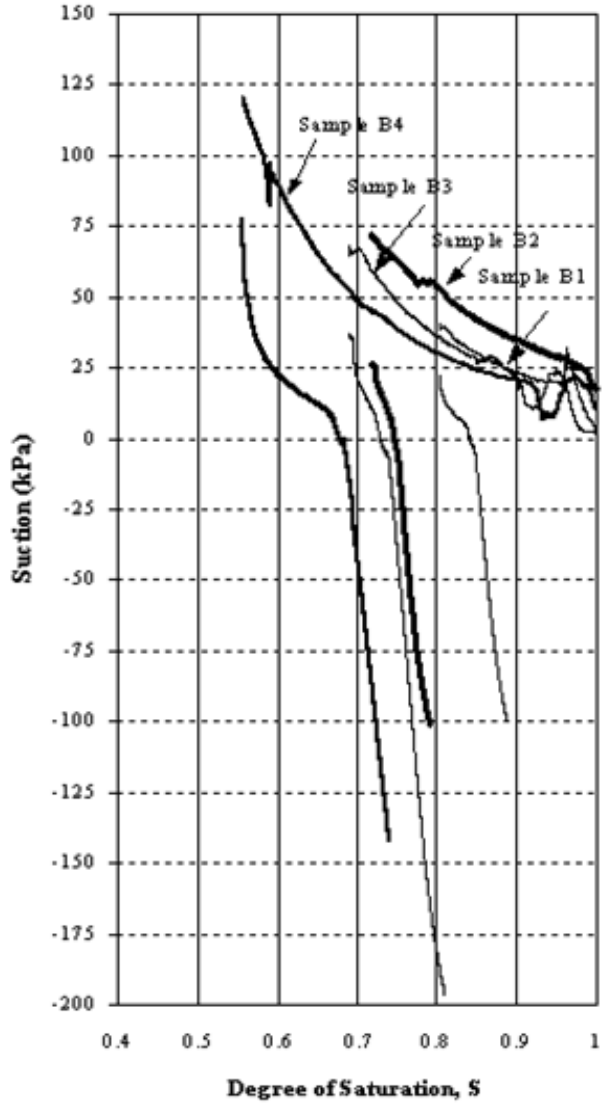

Figure 9 : WRC near saturation (*Bicalho, Znidarcic & Ho)*)

### 2.4 Correlation between WRC and some soil properties

Water transfer in porous media is obviously dependent on the pore size distribution of the medium and various early contributions (Childs and Collis-George 1950, Burdine 1953, Marshall 1958) proposed some statistical models to account for this aspect. As described by Fredlund et al. (1994), further investigation was developed in unsaturated soils based on the shape of the water retention curve by Kunze et al. (1968), Mualem (1976) and van Genuchten (1980). Some recent investigations in this field have been performed by Fredlund et al. (1994), Leong and Rahardjo (1997) and Huang et al. (1998), who examined in details the existing models and proposed some new expressions linking the permeability to the water retention curve.

#### 2.4.1 Shear strength

Some investigations have been carried out on the relationship between the WRC and the shear resistance of unsaturated soils, based on the idea that the WRC provided an insight on the nature of the inter-particle contacts and of the inter-granular stresses at various suctions (Fredlund et al. 1995, Öberg & Sällfors 1995, 1997, Vanapalli et al. 1996). In terms of soil microstructure, the physical idea appears to be best adapted for sands where punctual contacts may be observed between soil particles.

### 2.5 Mathematical expressions of the water retention curve

Various mathematical expressions of the WRC has been provided and synthesised in various general reports, books and papers (Alonso et al. 1987, Fredlund & Rahardjo 1993, Barbour 1998, Huang et al. 1998 among others). The detailed expressions can easily be found an are not given here. Recently, the relative merits of various existing mathematical expressions was investigated in details by Leong & Rahardjo (1997) and Sillers et al. (2001). These authors considered the well known expressions of Burdine (1953), Gardner (1956), Brooks & Corey (1964), Brutsaert (1966), Mualem (1976), van Genuchten (1980), Tani (1982), Mc Kee & Bumb (1984), Fredlund & Xing (1994) and evaluated them according to the number, the independence and the physical meaning of the parameters and the correctness of the prediction provided. Most models behaved satisfactorily and the authors expressed their preference to models able to reach a 1 000 MPa suction (pF7), corresponding to the dry state according to Corey (1957).

In this Conference, *Gersovitch & Sayão* discussed the validity of the WRC mathematical expressions of Haverkamp-Parlange, Gardner, van Genuchten and Fredlund-Xing as compared to experimental WRC from 11 Brazilian soils. They concluded that the three last expressions were satisfactory for Brazilian soils, an advantage of Gardner's expression being the smallest number of parameters required ($\alpha$ and $n$).

### 2.6 Recent trends

In recent years, considerable interest has been devoted to the water retention curves of engineered clay barriers made up of compacted swelling clays used in nuclear waste disposal at great depth, so as to predict the water infiltration rate in these barriers in contact with a saturated host rock. These samples are characterised by high plasticity indexes and high densities, with very high initial suction (several tens of MPa). Controlled relative humidity techniques are often used to determine the water retention curves, that sometimes present negligible hysteresis, due to high density and predominant physico-chemical effects. The effect of impeded swelling on the retention properties also appeared significant (Delage et al. 1998a & b).

The effects of the volume changes due to a constant applied stress on the determination of the water retention properties of a volcanic soil were investigated by Ng and Pang (2000), who also considered hysteresis effects.

Researches in this field also led to investigating the effects of temperature on the water retention and the mechanical properties of engineered barriers (Romero et al. 1995, 1998). In this Conference, an extension to the study of chemical effects on the water retention properties is presented by *Mata, Romero & Ledesma*.

## 3 SUCTION MEASUREMENTS

A complete description of the methods of measuring suctions was given by Ridley & Wray (1995) at the Paris UNSAT'95 Conference. Obviously, the Imperial College tensiometer (Ridley & Burland 1995) represented a significant advance towards the measurement of matrix suction up to 1.5 MPa, providing continuity between standard tensiometer (up to 80 kPa) and thermocouple psychrometer (up to 7 MPa). Further work on the IC tensiometer was carried out by Tarantino & Mongiovi (2001) and *Tarantino & Mongiovi* (this Conference). The IC tensiometer was also used by *Villar & De Campos*, *Marinho, Kuwagima, Standing & Fulton*.

Besides the IC tensiometer, other techniques used in this Conference to measure suction in the laboratory or in-situ comprise the standard tensiometer, the filter paper, a granular matrix sensor, a sponge (*Bertolino, Souza Fernandes, Rangel, de Campos & Shock ; Mahler, Mendes, Souza & Fernandes ; Montrasio; Villar et al.*). *Shuai, Clements, Ryland & Fredlund* provide a further insight on the effects of temperature, freeze and pH on the response of thermal sensors.

In this Conference, an increasing tendency to measure suction changes in the triaxial apparatus has been observed. Figure 10 (*Colmenares & Ridley*) shows the results of unconfined compression tests carried out at various constant water contents on sample of a reconstituted clayey silt that were dried during various periods of time. Apparently, suction decrease occur in more fragile samples at suctions higher than 200 kPa, probably in relation with larger elastic zones.

*Becker & Meißner* used simultaneously standard tensiometers and the axis translation technique and they monitored soil suction between 85 and 285 kPa. *Kawai, Weichuan & Ogawa* monitored suction changes during isotropic tests under a constant (larger) air pressure by measuring the changes in (smaller) positive water pressure.

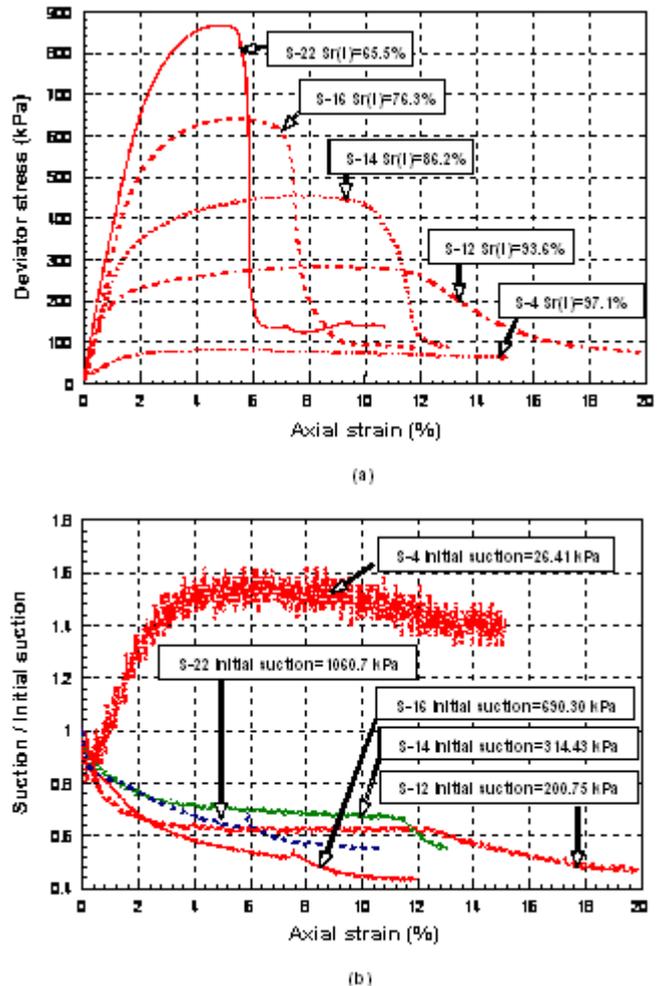

Figure 10 : Stress strain and suction-strain relationship in a constant water content test (*Colmenares & Ridley*)

## 4 THE CONTROL OF SUCTION IN SOIL MECHANICS APPARATUSES

### 4.1 *Introduction*

As seen in Figure 1, standard geotechnical apparatuses have been adapted so as to control the suction inside the sample. Most systems are based on the axis translation method, imposing a positive air pressure $u_a$ inside the sample and controlling the water pressure $u_w$ (generally equal to atmospheric pressure, but sometimes controlled at a given value). Technically, the adaptation of this technique in the triaxial apparatus is not too complicated, as seen in the Figure. The lower porous stone is replaced by a high air entry value ceramic porous stone, and air pressure is injected through the upper porous stone.

Semi-permeable membranes are sometimes used instead of ceramic porous stones.

Other well known applications of the axis translation method to the oedometer and to the direct shear box are presented in Figure 11 (Escario 1967, Escario & Saez 1973, Escario 1980). For safety reasons, particular attention has to be paid to the mechanical resistance of the cells which contain air pressures, as can be seen in the Figure. Air tightness is to be ensured at the contacts of the pistons that are used to apply vertical stresses and related friction can affect the exact value of the stress applied on the sample. Friction may also affect the rod that allows for the mutual displacement of the two half-boxes of the shear box. The confining cell of the shear box has to be large enough to allow for shear displacement. These systems have been extended to suctions as high as 12 MPa (Escario & Juca 1989). Suction controlled shear boxes were also developed by Gan et al. (1988).

In a direct shear test, the sheared zone is very localised and thin, as compared to the mass of the unsaturated sample. When analysing the status of suction in an unsaturated sample submitted to direct shear stress under a slow shearing rate, it seems highly probable that a regulation of the suction in the sheared zone could be done by the mass of soil, since the volume of water to be exchanged to maintain suction equilibrium in the whole sample is small. Provided that the shearing rate is low enough, it seems highly probable that a slow constant water content direct shear test would also ensure a condition of constant suction. Experimental checking of this thinking would be welcome. In this Conference, a direct shear box was used by *Bilotta & Foresta*. *Perez-Garcia & Cortez-Ochoa* developed a suction controlled oedometer similar to that of Lloret (1982).

Other techniques of controlling suction have been applied to geotechnical testing. The osmotic technique was adapted to the oedometer by Kassiff & Ben Shalom (1971). Figure 12 presents the system of Kassiff & Ben Shalom completed by a system allowing PEG circulation inside a closed circuit proposed by Delage et al. (1992). The capillary tube placed in the closed bottle that contains the solution allows the monitoring of water exchanges during the tests. An alternative method of controlling water exchanges by placing the bottle on a balance has been proposed by Dineen & Burland (1995). As seen previously, the maximum possible suction attainable with the osmotic technique has been extended up to 10 MPa by increasing the PEG concentration (Figure 2). The osmotic system has also been adapted to triaxial testing (Delage et al. 1987, Cui & Delage 1995).

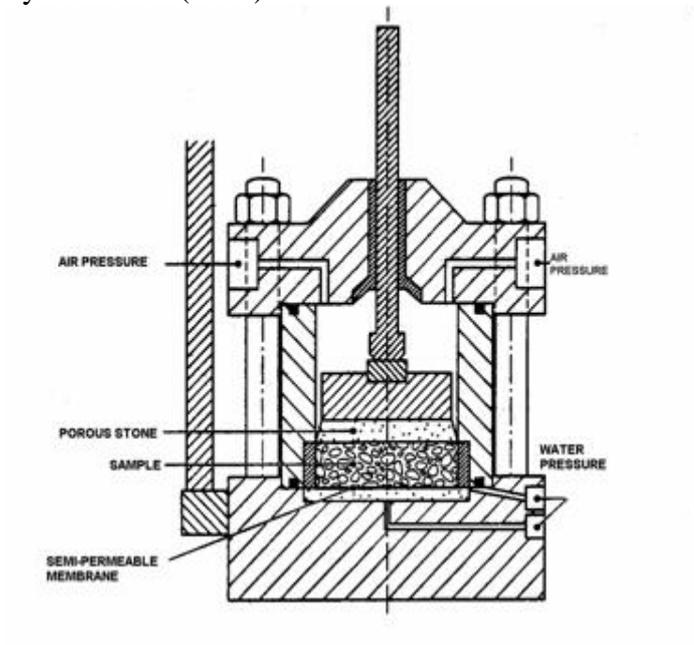

a)

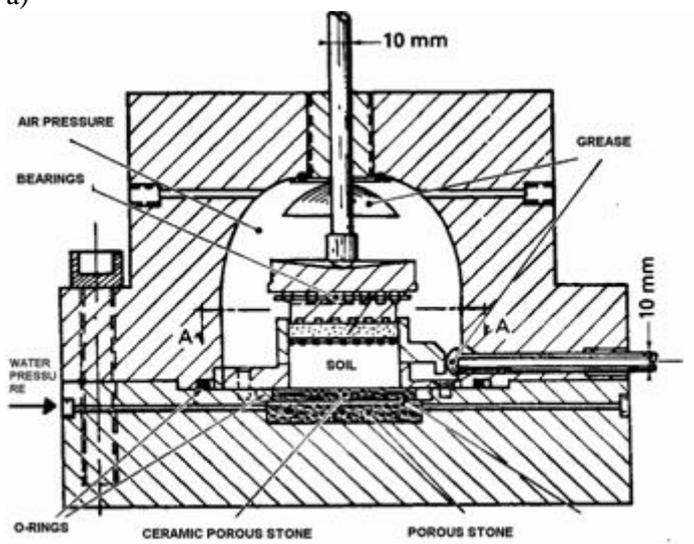

b)

Figure 11 : a) suction controlled oedometer (Escario & Saez 1973 – b) suction controlled shear box (Escario 1980)

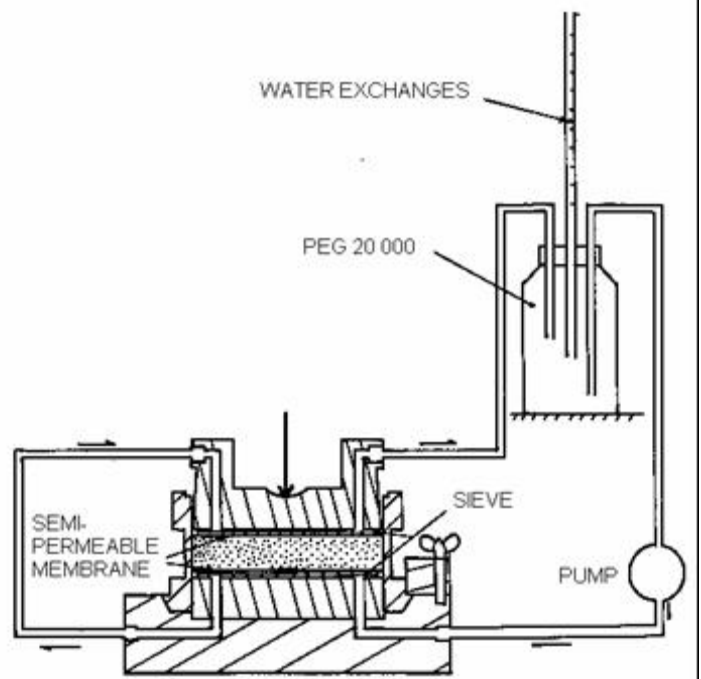

Figure 12 : Osmotic oedometer (after Kassiff & Ben Shalom 1971 and Delage et al. 1992)

The vapour equilibrium technique, based on the control of the relative humidity of the atmosphere surrounding the sample (using saturated salts or sulphuric acid at various concentrations) has been applied for first time to the oedometer by Esteban (1990). Figure 13 (Oteo-Mazo et al. 1995) shows the principle of the apparatus. Since water transfer occur in the vapour phase, this system controls the total suction.

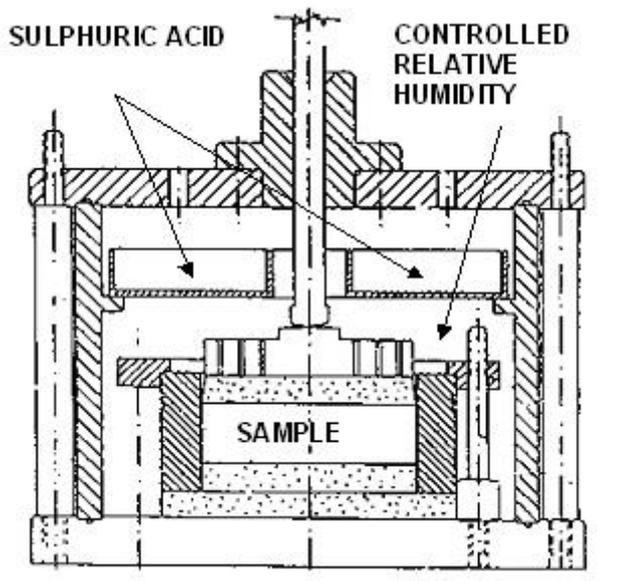

Figure 13 : Oedometer with controlled relative humidity (Esteban 1990, Oteo-Mazo et al. 1995)

The time necessary to reach equilibrium is quite long, due to the very slow rate of vapour exchange between the saline solution and the sample. This technique allows the application of very large suctions (a saturated solution of KOH, with a 9% relative humidity, imposes a 332 MPa suction). It is suitable for tests on engineered barriers, where suction values larger than 50 MPa are currently encountered. The use of this system is increasing in research devoted to nuclear waste disposal (Al Mukhtar et al. 1993, Bernier et al. 1997).

4.2 *Triaxial apparatus*

In the axis translation system, the necessity of increasing simultaneously the confining pressure $\sigma_3$ and the air pressure $u_a$ so as to apply the initial net stress $\sigma_3 - u_a$ results in a limitation of the suction $u_a - u_w$ applied, due to the limitations in pressure of the standard triaxial cells generally used (up to 2 MPa). In the literature, the suction applied in the triaxial using this technique are generally smaller than 400-500 kPa. As commented before, higher suctions (up to 12 MPa) have been achieved in the shear box and should also be achieved in the triaxial apparatus using high pressures cells. Other problems encountered during long term tests (1 week or more) are related to the diffusion of air through the membrane and in the water of the porous stone.

An interesting application of the vapour equilibrium technique, presented in Figure 14, has been recently published by Blatz and Graham (2000).

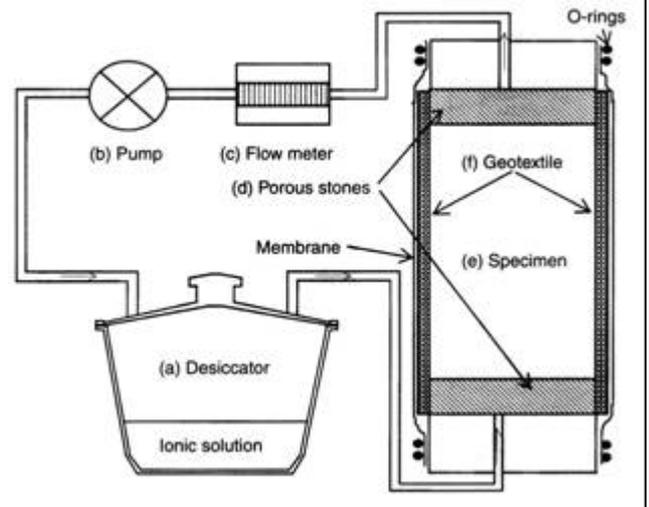

Figure 14 : Triaxial with vapour controlled system (Blatz & Graham 2000)

In this system, water vapour is circulated in a porous geosynthetic placed around the triaxial sample, between the sample and the membrane. As shown in the Figure, the geosynthetic is in contact with the upper and lower porous stones. This device reduces the drainage length that characterises the rate of vapour exchange and the equilibrium period. The triaxial system also comprises thermocouple psychrometers (a top cap and an internal psychrometer) that give a direct measurement of the suction changes during shearing.

In this Conference, a significant number of contributions are based on the use of advanced suction controlled triaxial apparatuses, based on the axis translation method (*Barrera, Romero, Lloret & Vaunat ; Becker & Meißner ; Carvalho, Lloret & Josa ; Nishigata, Fleureau, Hadiwardoyo, Dufour-Laridan, Langlois & Gomez-Correia ; Kawai, Weichuan & Ogawa ; Nishida & Araki ; Rifa'i, Laloui & Vulliet*).

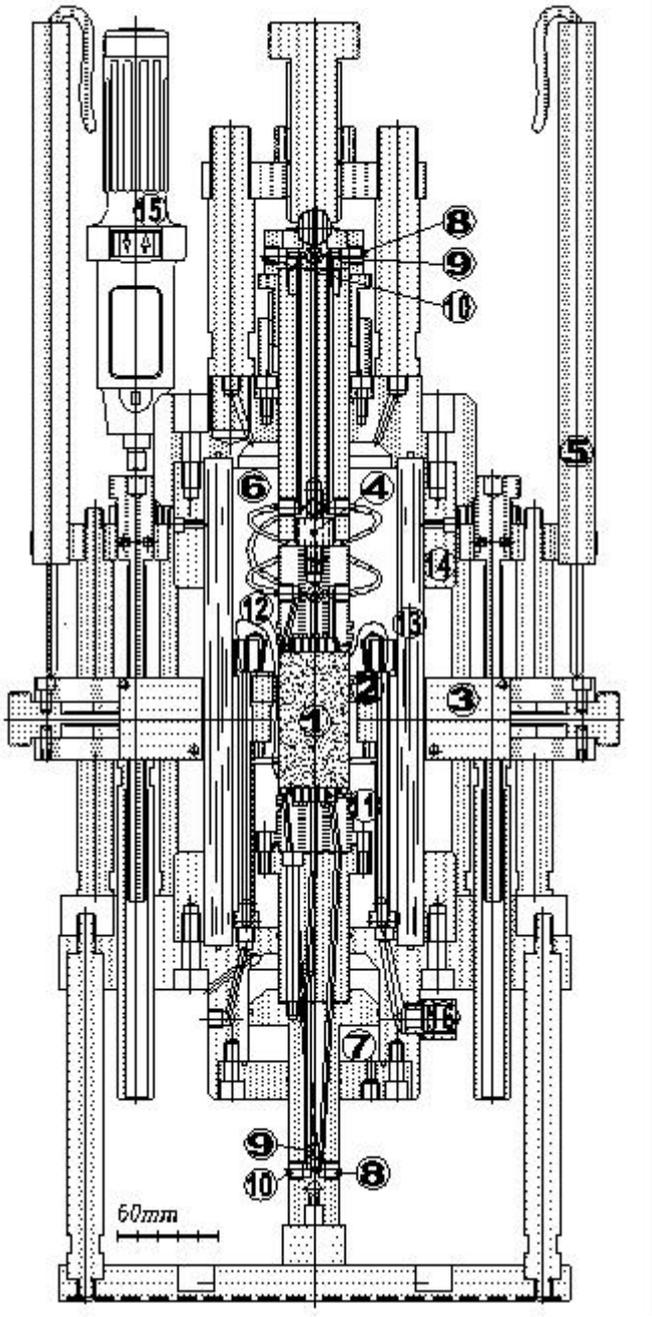

Figure 15 : Barcelona advanced triaxial cell (Romero et al. 1997, *Barrera et al.*)

Figure 15 presents an advanced suction controlled triaxial cell developed in Barcelona (Romero et al. 1997, *Barrera et al.*) where the exact shape of the sample is monitored during shearing by means of a mobile electro-optical laser system mounted outside the chamber (see also Figure 17).

4.2.1 *Testing rates*

Controlled suction triaxial tests in unsaturated soils are comparable to drained saturated triaxial tests. At any time, the system that controls suction has to provide or extract the amount of water necessary to counterbalance the effects of the triaxial compression. Experience (see Rampino et al. 1999) shows that water exchanges related to the control of suction are mainly controlled by the volume changes : shear contraction imposes water drainage, whereas shear dilatancy imposes water extraction. As in saturated drained tests, the rate of shearing is depending on the suction changes induced by the volume changes, on the drainage length of the sample and on the water permeability of the sample, that decreases with increased plasticity index. A further condition is imposed in unsaturated soils testing by the low permeability and the thickness (i.e. the impedance) of the HAEV porous stone or semi-permeable membrane used. This problem has been treated by Ho & Fredlund (1982b) based on the calculations made by Gibson & Henkel (1954) for saturated soils. In a saturated sample, the pore pressure generation induced by the volume changes during shearing at a given strain rate is easy to estimate, whereas the generation of suction during the shearing of an unsaturated soil at a given strain rate is much more difficult.

Obviously, few experimental data are presently available in this regard, and the topic needs further investigation, since it is also of interest for the understanding of many engineering problems. Ho & Fredlund (1982b) provided an estimation of the strain rate close to the rate of saturated drained tests, i.e. close to 1μm per minute. A similar value was obtained with the osmotic technique considering the impedance of a semi-permeable membrane (Delage et al. 1987).

The drainage length of the Bishop & Donald's system (Figure 1), that is used in most of triaxial testing investigations, is equal to the length of the sample. This length has been reduced to half the sample length by Maâtouk et al. (1995) who injected the air through a thin tubing connected at the middle of the sample through the cylindrical membrane. An alternative solution, proposed by Romero et al. (1997), consists in controlling air and water pressures at both ends of the sample, by using a peripheral annular coarse porous stone for air, and a 3 mm diameter HAEV porous stone. Data obtained with this system, that will be further described, are presented in this Conference in *Barrera, Romero, Lloret & Vaunat*. In the osmotic triaxial apparatus, the placement of semi-permeable membranes at top and bottom of the sample also provides a H/2 value of the drainage length.

The rate of shearing used by various authors is in this order of magnitude, as can be seen in Table 1.

Table 1 : Rate of shearing used in controlled suction triaxial testing

| Author | Soil | Drainage Length | Shearing Rate (μm/mn) |
|---|---|---|---|
| Bishop & Donald (1961) | Brahead Silt | H | 2.13 |
| Gulhati & Satija (1981) | Dhanauri Clay $I_p = 24$ | H | 6 |
| Ho & Fredlund (1982) | Silty sand, sandy silt | H | 1;43 |
| Delage et al. (1987) | Jossigny LP silt $I_p = 19$ | H/2 | 1 |
| Romero et al (1997), *Barrera et al.* | Barcelona LP silt $I_p = 16$ | H/2 | 1 |
| Wheeler & Sivakumar (1995) | Speswhite Kaolin $I_p =$ | H | 1.4 |
| Maâtouk et al. (1995) | LP loess $I_p = 7$ | H/2 | |
| Rampino et al. (1999) | Silty sand $I_p = 13$ | H | 1.26 |
| Laloui et al. 1997, Geiser 1999 | Sion sandy silt $I_p = 8$ | H | 1.5 |
| Rampino CRS | Silty sand $I_p = 13$ | H | 5 |

4.2.2 *Volume changes monitoring*

The method proposed by Bishop & Donald (1961, Figure 1) has been widely adopted afterwards, because the simpler alternative technique consisting in monitoring directly the exchanges of the water contained in the cell is known to present various disadvantages (Geiser et al. 2000). Automatic monitoring of the mercury level was proposed by Josa et al. (1987) by using a metal ring floating on the mercury surface, the vertical displacement of the ring being monitored by a submerged LVDT transducer. Progressively, the use of mercury was progressively abandoned for safety reasons. Cui & Delage (1996) replaced the mercury by water and used two confining fluids : the cell was filled up by water up to a level located just below the top of the glass cylinder, the remaining top volume of the cell being filled with air under pressure. A thin film of silicone oil was placed on the surface of water to prevent evaporation. The level of the silicone oil was followed optically, and the precision of the system was enhanced by using a large diameter piston (close to the sample diameter), so as to minimise the area comprised between the glass cylinder and the piston and optimise the precision by increasing the vertical movements. Rampino et al. (1999) also used air as confining fluid and developed a specially designed cylinder to improve the precision. They monitored the change in the water level with a differential pressure transducer. Aversa & Nicotera (2002) described in detail a triaxial (and oedometer) cell based on a similar principle. *Nishigata, Nishida & Araki* also used air in the cell and water in the surrounding cylinder and used a gap sensor to monitor the vertical movements of the water. *Carvalho, Lloret & Josa* improved the system of Josa et al. (1987) by using a proximity sensor to monitor the displacements of a ring floating on the fluid (silicon oil) contained in the inner cylinder.

Wheeler (1988) modified Bishop and Donald's system by using a double cell triaxial cell, as shown in Figure 16. Most disadvantages of the direct cell water volume measurements are eliminated because the inner cell does not experience any stress changes during increases in confining pressures. The advantages and draw-backs of this technique are discussed in details by the author, the main problem being the water adsorption by the perspex cylinder. This can be solved by using a glass inner cell.

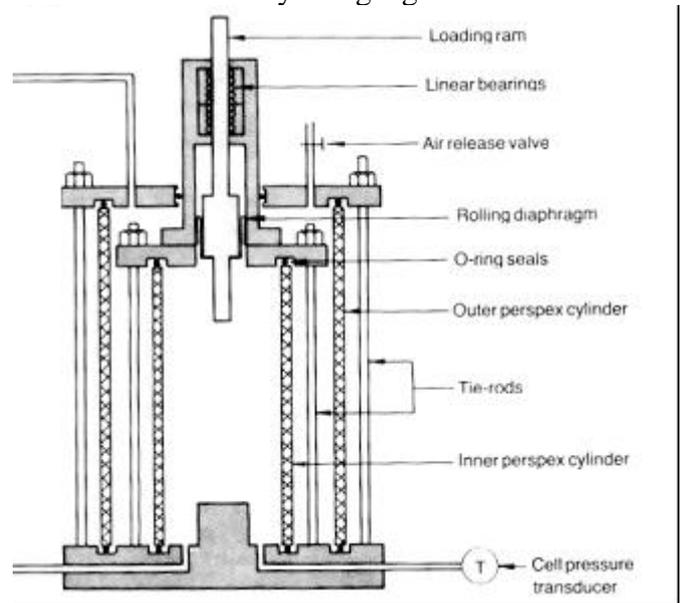

Figure 16 : Double wall triaxial cell (Wheeler 1988)

Another methods for monitoring volume changes are based on the use of local measurements, as described by *Elkady, Houston & Houston* : LVDT, Hall effects transducers (Clayton & Khatrus 1986, Maâtouk et al. 1996, Geiser et al. 2000, *Carvalho et al., Becker & Meißner, Fleureau et al., Kawai, et al.*), strain gauges (Lo Presti et al. 1995, Kolymbas & Wu 1989), proximity sensors (Cole 1978, Khan & Hoag 1979, Drumright 1989) and opto-electronic sensors (Baumgartl et al. 1995). Laser techniques were also developed by Romero et al. (1997), providing a complete description of the profile of the sample during shearing.

A comprehensive description and evaluation of the method of controlling volume changes in triaxial testing of unsaturated soil was provided by Geiser et al. (2000), who commented the respective advantages and disadvantages and gave the absolute errors of various techniques that they tried. The paper also provides the absolute precision of the methods, reported in Table 1. As compared to the techniques previously described, a novel approach based on image processing (Macari et al. 1997) is also introduced and evaluated.

Table 2 : Characteristics of existing methods of measuring volume changes in unsaturated triaxal testing (after Geiser et al. 2000)

| Method | Absolute error<br>$\alpha$ : volume change<br>$\beta$ : volumetric strain |
|---|---|
| Standard triaxial cell | Geiser :<br>$\alpha = \pm 0.45$ cm$^3$<br>$\beta = \pm 2.2\ 10^{-3}$ |
| Cell with inner open cylinder (Bishop & Donald) | Geiser :<br>$\alpha = \pm 0.21$ cm$^3$<br>$\beta = \pm 10^{-3}$<br>Bishop (100 cm$^3$):<br>$\alpha = \pm 0.21$ cm$^3$<br>$\beta = \pm 0.8\ 10^{-3}$ |
| Double walled cell (Wheeler 1988) | Sivakumar 1993<br>$\alpha = \pm 0.6$ to $1.02$ cm$^3$<br>$\beta = \pm 6\ 10^{-3}$ to $10^{-2}$ |
| Air filled controller | Geiser :<br>$\alpha = \pm 2.2$ cm$^3$<br>$\beta = \pm 1.1\ 10^{-3}$<br>(possible leakage) |
| Radial local measurement | $\alpha =$<br>$\beta =$ |
| Laser | Romero et al. 1997<br>$\beta = \pm 0.7\ 10^{-4}$ |
| Image processing | Geiser :<br>$\alpha = \pm 0.25$ cm$^3$<br>$\beta = \pm 6\ 10^{-3}$ |

In this Conference, the same group (*Rifa'i, Laloui & Vulliet*) compared image processing data to that provided by the volume change of the confining water in a standard triaxial cell, and found reasonable agreement. Image processing was also used by *Elkady, Houston & Houston* who commented on the magnification corrections due to water and cell during a test performed on a rubber specimen. The pattern of axial strains that they obtained is somewhat similar to that obtained by using the laser technique by *Barrera, Romero, Lloret & Vaunat* and presented in Figure 17. Local and global lateral strains are compared, evidencing the progressive barrelling of the sample.

This reveals a major problem encountered in the complete determination of the state of an unsaturated soil, when the degree of saturation can only be defined in a global manner, whereas strains are not homogeneous. In other words, local radial and axial measurements, known as being more satisfactory because not altered by the friction effects on top and bottom of the sample, cannot be related to a global value of the degree of saturation. In this regard, novel techniques of measuring the degree of saturation such as the X Ray tomography measurements carried out by Wong & Wibowo (2000) in a sand is certainly of interest. In this Conference, image analysis is used for the determination of the degree of saturation of a sand by *Sharma, Mohamed & Lewis*, based on the changes in colour that occurs when the water content changes.

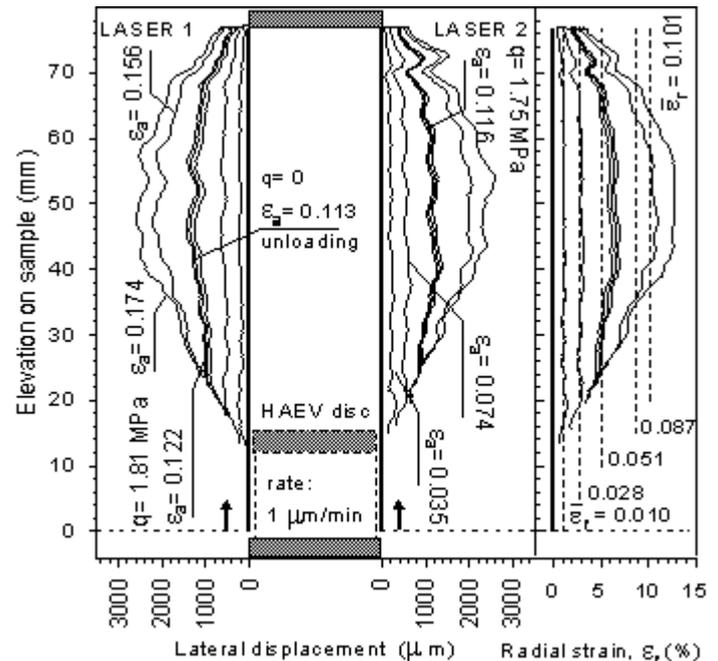

Figure 17 : Progressive development of local and global strains during shearing (*Barrera et al.*)

## 5 OBSERVED EXPERIMENTAL BEHAVIOUR OF UNSATURATED SOILS

The triaxial apparatus provides, in an axisymetric state of stress ($\sigma_2 = \sigma_3$), a complete set of data characterising the constitutive behaviour of soils, evidencing for instance the combined effects of volume change and shear that are typical of plastic behaviour.

### 5.1 *Volume changes*

Earlier investigations carried out in the framework of two independent stress variables (Coleman 1962, Matyas & Radhakrishna 1968, Fredlund & Morgenstern 1977, Tarantino et al. 2000) concerned volume changes properties under either controlled suction compression tests (using either an isotropic cell - Matyas & Radhakrishna 1968 or an oedometer - Barden et al. 1969) or volume change under a constant stress and changing suction. The concept of state surfaces evidenced by Matyas & Radhakrishna to describe the changes in void ratio and degree of saturation (under a limiting condition of increased degree of saturation) actually constituted the first proposed behaviour law for volume changes in unsaturated soils.

5.1.1 *Compression tests*

Figure 18 shows a recent and complete set of data of constant suction isotropic compression tests carried out on a compacted Metramo silty sand ($I_p = 13$, $w = 9.8\%$, $\gamma_d = 19.7\ kN/m^3$) by Rampino, Mancuso & Vinale (1999).

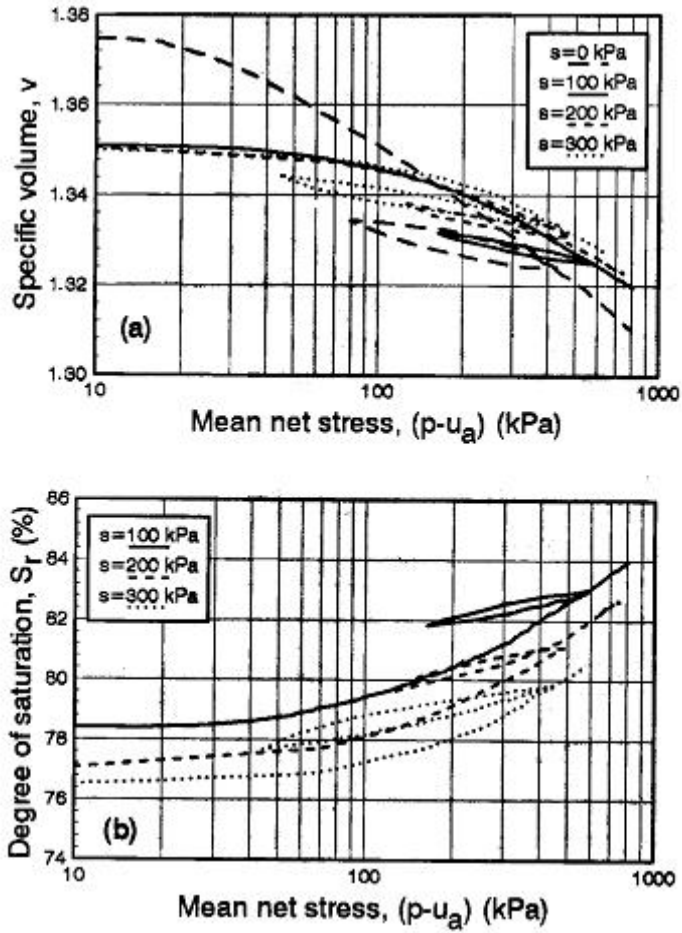

Figure 18 : Constant suction isotropic compression tests carried out on samples of compacted Metramo silty sand (Rampino et al. 1999)

The as-compacted suction being 800 kPa, the samples tested are located on the main wetting path of the WRC at $s = 300$, 200 and 100 kPa respectively. The data confirm the general findings synthesised for first time by Alonso et al. 1987 : an increased suction produces i) a stiffening of the soil with smaller plastic compressibility indexes $\lambda$ and ii) an increase of the yield stress, leading to the definition of the locus of the yield points in the suction/net mean stress plane named LC (see also Alonso et al. 1990). The Figure also shows that the elastic modulus $\kappa$ determined (in a Cam-clay $v/ln(p - u_a)$ diagram) by running cycles in stress is not suction dependent. The changes in degree of saturation also illustrate an irreversible response, confirming the idea of Wheeler (1997) who suggested to adopt an elasto-plastic approach for the changes in water content of unsaturated soils presenting an elasto-plastic strain-stress behaviour.

Toll (2000) also proposed a simple quantitative model to account for these effects. It is interesting to note that the slopes and yield stresses defined in Figure 18 seems to be independent of the suction. This aspect, which obviously needs further confirmation, should allow for interesting simplifications if confirmed.

5.1.2 *Collapse tests*

An important feature of the volume change behaviour of unsaturated soils is the phenomenon of collapse observed when the soil is wetted under load. Many publications deal with this aspect (see for instance Jennings & Knight 1957, Northey 1969, Sultan 1969, Dudley 1970, Barden et al. 1973, Clemence & Finbarr 1981, Yudhbir 1982, Mitchell 1993, Derbyshire et al. 1995, Pereira & Fredlund 2000) which can be modelled using the state surface of Matyas & Radhakrishna (1968). Collapsible soils are most often low plastic loose unsaturated soils and the collapse corresponds to the collapse of an initially metastable fabric.

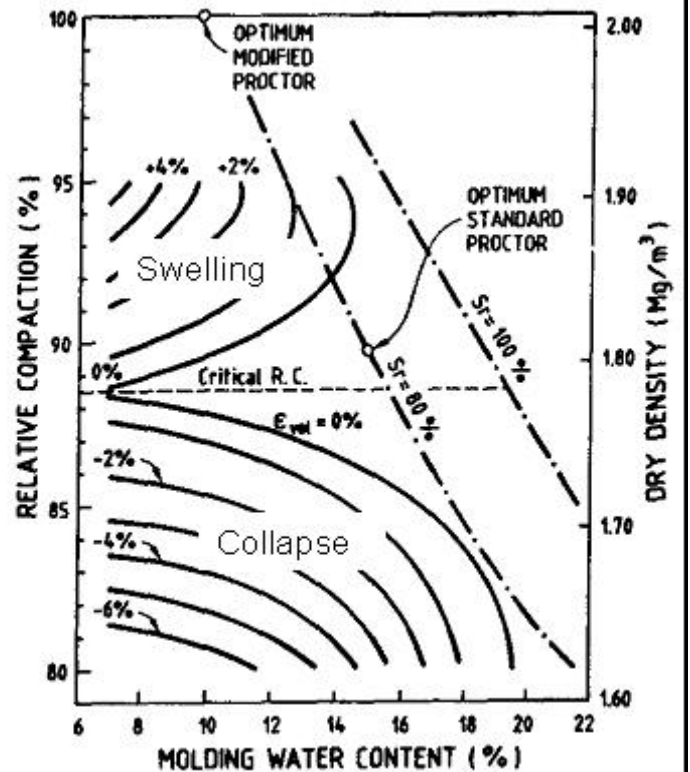

Figure 19 : Collapse behaviour of a compacted clayey sand (Lawton et al. 1989 in Gens 1995))

The level of collapse depends upon the density and initial water content of the soil. A general pattern on collapse properties of a slightly expansive compacted clayey sand is provided in Figure 19, where the tendency of collapsing or swelling under wetting as a function of water content and density is plotted using contours of equal strain (Lawton, Fragaszy & Hardcastle 1989).

Further insight was also provided by studying the effect of the stress anisotropy on the magnitude of collapse (Lawton et al. 1991).

Figure 20 presents data from Kato & Kawai (2000) where samples have been wetted from $s = 245$ kPa down to 0 under various deviator stresses. Results are presented in terms of volumetric and shear strains, they show the influence of the $q/p$ stress ratio on the magnitude of collapse. The same authors propose a model to predict this behaviour. This kind of investigation can be related with the contribution of *Schreiner & Okonta*, which considered the shear strain induced by wetting at the pile-soil interface.

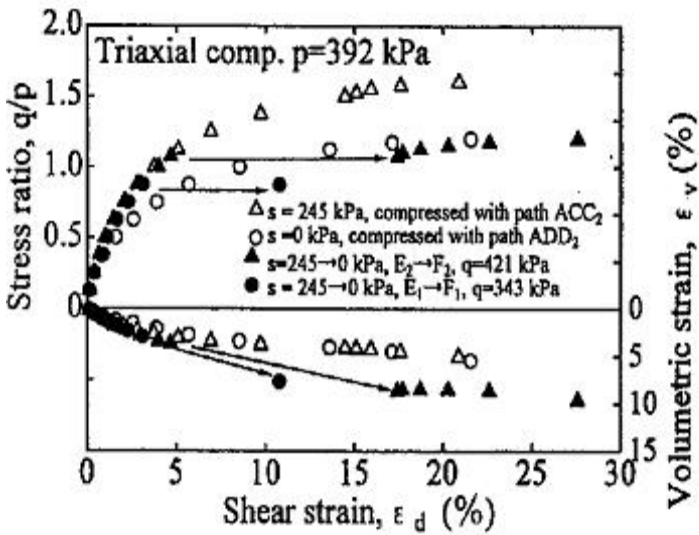

Figure 20 : Effect of strain anisotropy on the magnitude of collapse (Kato & Kawai 2000)

## 5.2 Elasto-plastic behaviour

Actually, the volumetric behaviour observed in Figure 18 exhibited an elasto-plastic behaviour with plastic strain occurring above the yield stress. A plastic behaviour comprising a deviatoric effect can also be observed in the data from controlled suction triaxial presented in Figure 21 (Rampino et al. 1999), where the deviator stress $q$, the volume changes $\delta v$ and the changes in water contents $\delta S_r$ are plotted as a function of the axial strain $\varepsilon_a$. In these samples that follow the wetting path from an initial suction of 800 kPa, observation at small axial strains (< 1%) shows that the decreased suction induces a decrease in the elastic shear modulus.

Hence, suction, that had no significant effect on elastic volume changes (Figure 18), seems to have some effect on the elastic shear modulus. This is confirmed by results from Marinho et al. (1995) shown in Figure 22, that show the changes in elastic shear modulus at small strain as a function of suction investigated on samples of London clay ($I_p = 53$) compacted at various densities, using bender elements. In all samples, the modulus first increases with increased suction, followed by either a maximum or a stabilisation at higher suctions.

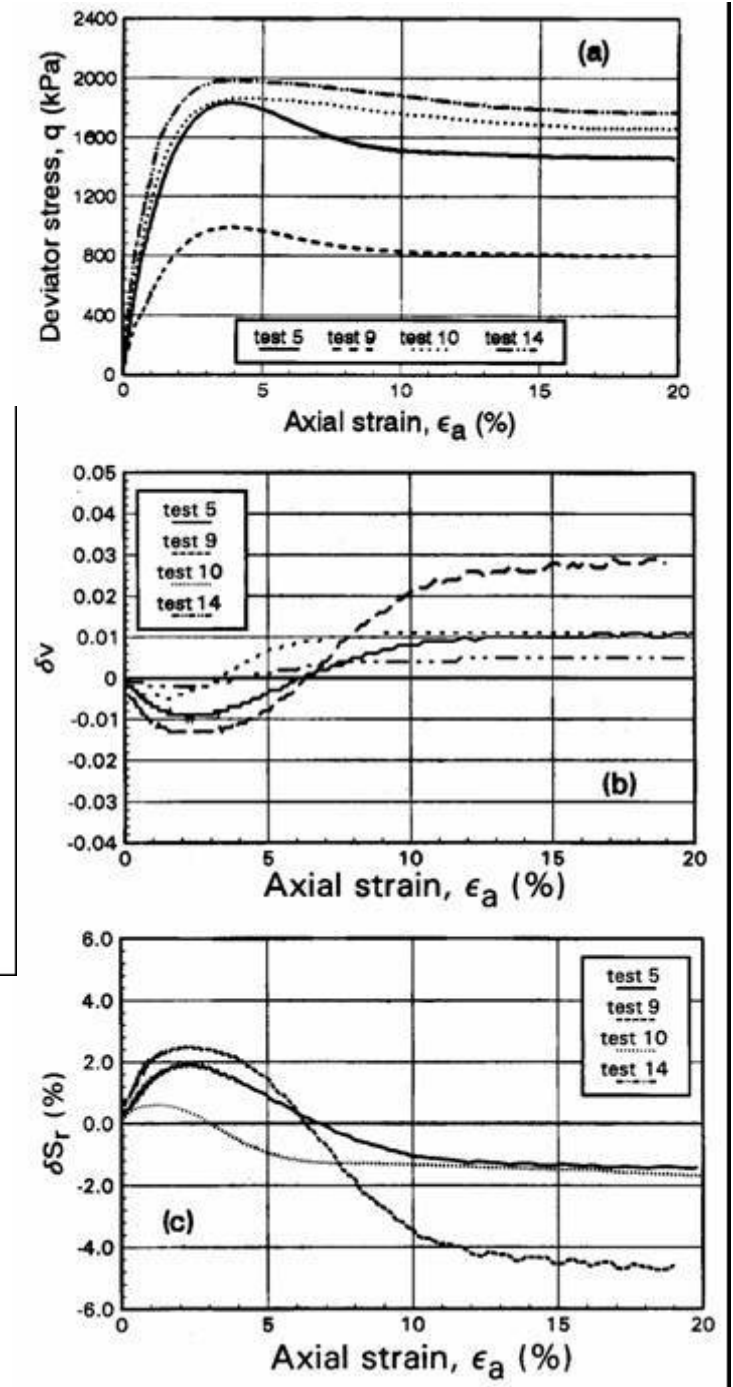

Figure 21 : Stress-strain curves obtained from controlled suction triaxial tests at $\sigma_3 - u_a = 400$ kPa (Rampino et al. 1999)

The elastic behaviour at small strains is investigated in this Conference by *Fleureau et al.* based on resonant column tests that are compared with measurements using a high precision triaxial device. *di Mariano, Vaunat & Romero* also study the elastic volumetric behaviour of a silt in a suction controlled oedometer. Based on the consequent existing literature data available on the behaviour of saturated soils in the elastic domain, this point obviously deserves further investigation.

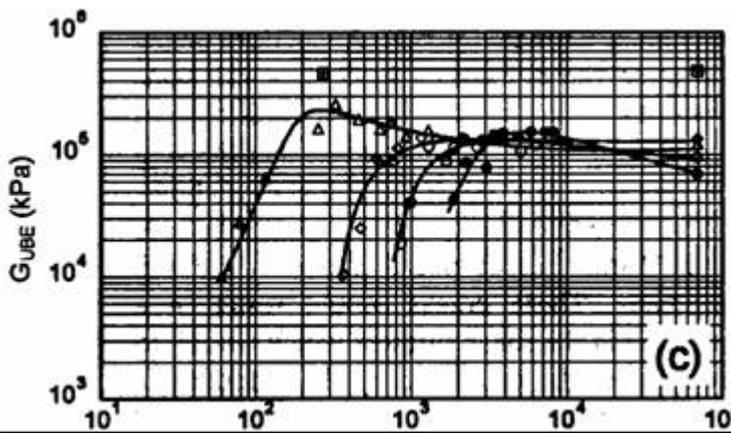

Figure 22 : Changes in the shear modulus with suction (Marinho & Chandler 1995).

Figure 21 also shows how plastic dilatancy progressively compensates the elastic compression strains so as to reach a point of phase change where the behaviour changes from contractancy to dilatancy (between 1.5 and 2.5% axial strain), just before reaching failure. The yield point is however not really apparent on the $q/\varepsilon_a$ shear curve. Observation of the response in degree of saturation shows that the water exchanges are mainly due to volume changes. Their absolute changes are larger at higher suction, i.e. at smaller degree of saturation.

In the framework of elasto-plasticity and critical state concepts, the shape of the elastic zone (in a $q/p'$ plane in saturated soils) was investigated in a $q / p - u_a / u_a – u_w$ space, within a framework defined by Alonso et al. 1990 by Cui (1993), Cui et al. (1995), Zakaria (1995), Zakaria et al. (1995) and Cui & Delage 1996. Figure 23 shows the elastic zones and plastic strains observed in an isotropically compacted soil (Wheeler & Karube 1995) and an anisotropically compacted silt (Cui & Delage 1996). Figure 23a shows a symetrical yield locus with associated plastic strain vectors, whereas Figure 23b shows an elliptical yield curve inclined along the $K_0$ axis, typical of the structure anisotropy induced by 1D compaction of the sample, with non-associated plastic strains. Figure 23a evidences the isotropic hardening induced by an increased suction also observed in the anisotropically compacted sample. In Figure 23b a slight reduction of the inclination of the yield curve due to the application of an isotropic stress on the anisotropic sample is noticeable in the extrapolated yield curve at 600 kPa, resulting in a combined stress hardening. These aspects are considered by *Romero, Barrera, Lloret & Gens* in this Conference.

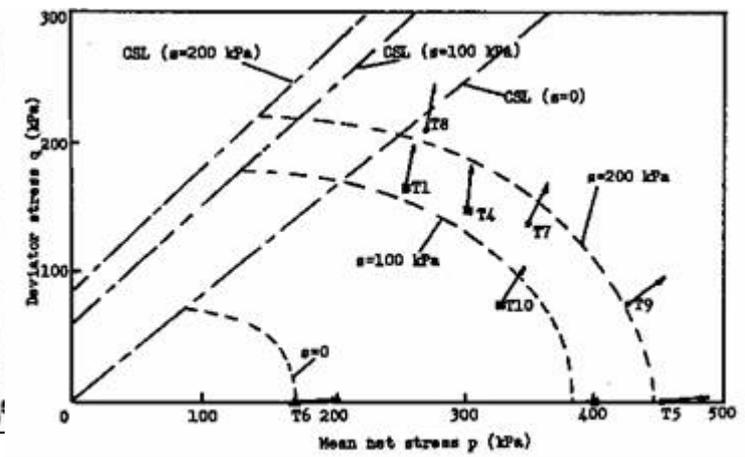

a)

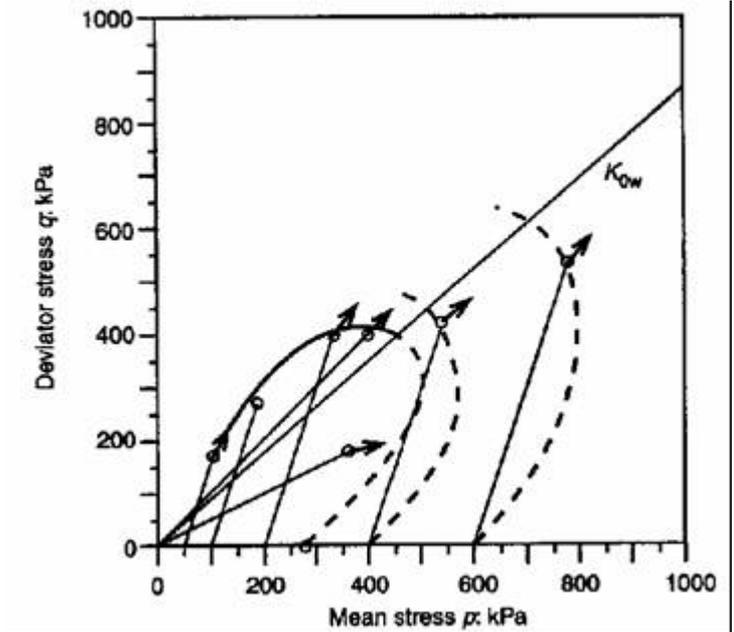

b)

Figure 23 : Elasto-plastic behaviour of compacted soils :
a) isotropically compacted soil (Wheeler & Karube 1995)
b) anisotropically compacted silt (Cui & Delage 1996)

### 5.3 *Shear behaviour*

For engineering purposes, the characteristics at failure (cohesion $c$ and friction angle $\varphi$) are of interest in the design of structures. They were investigated in the past by various authors, often based on direct shear box data (Escario & Saez 1986, Gan et al. 1988, De Campos & Carillo 1995, Fredlund et al. 1995). The most commonly used shear strength criterion was proposed by Fredlund et al. (1978), who introduced a $\varphi^b$ angle as follows :

$$\tau = c' + (\sigma - u_a) \tan \varphi' + (u_a - u_w) \tan \varphi^\beta \qquad (1)$$

Further investigations showed that $\varphi^\beta$ was not a constant but was decreasing with increased suction, up to a constant (or decreasing) value at high suction (Escario & Saez 1986, Fredlund et al. 1987, Escario 1988, Gan et al. 1988, Escario & Juca 1989).

Considering the following expression of Bishop (1959) "effective" stress :

$$\sigma' = [(\sigma - u_a) + \chi(u_a - u_w)] \quad (2)$$

equation (1) can be rewritten as :

$$\tau = c' + tan\,\varphi' \left[(\sigma - u_a) + \frac{tan\,\varphi^b}{tan\,\varphi'}(u_a - u_w)\right] \quad (3)$$

Considering the following expression of Bishop's $\chi$ parameter :

$$\chi = \frac{tan\,\varphi^b}{tan\,\varphi'} \quad (4)$$

the failure criterion (1) can be expressed as a function of Bishop's stress. Actually, the description of failure properties was one of the applications of Bishop's "effective" stress approach (Bishop & Blight 1963), and further research was carried out in this direction (Fleureau et al. 1995, Fleureau & Indarto 1995, Oberg & Sallfors 1997, *Fleureau, Hadiwardoyo, Dufour-Laridan, Langlois & Gomes-Correia*). In the same context, Khalili & Khabbaz (1998), based on experimental results from 13 different soils found in the literature, proposed the following expression of Bishop's $\chi$ parameter to describe the shear resistance of unsaturated soils :

$$\chi = \left[\frac{(u_a - u_w)}{(u_a - u_w)_b}\right]^{-0.55} \quad (5)$$

were $(u_a - u_w)_b$ is the air entry value of the sample.

More recently, Vanapalli & Fredlund (2000) discussed on various soils the relative performances of various criteria found in the literature in the shear strength/suction plane (Figure 24). The criteria provided by Vanapalli et al. 1996, Fredlund et al. 1996 and Vanapalli et al. (1998) all come from correlation with the soil water characteristic curve.

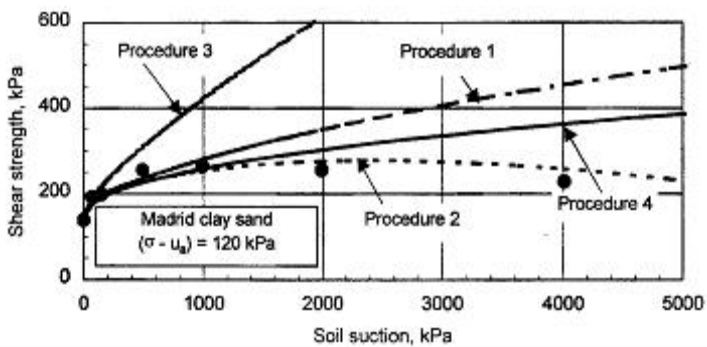

Figure 24 : Comparison of various shear strength criteria (Vanapalli & Fredlund 2000) : proc. 1, 2 and 3 from SWCC curves, corresponding to 1 : Vanapalli et al. (1996) and Fredlund et al. (1996) ; 2 : Vanapalli et al. (1998) ; 3 : Oberg & Sallfors (1997) ; 4 : Khalili & Khabbaj (1998).

Equation (1) is implicitly based on the hypothesis that the friction angle $\varphi'$ is independent of the suction. This has been checked in many cases, but it is not always true. Delage & Graham (1995) gathered various experimental data and observed that (Figure 25) whereas the cohesion was still increasing with suction, the friction angle could increase (Escario & Saez 1986 on a plastic clay, Drumright & Nelson 1995 on a copper tailings sand) or decrease (Delage et al. 1987, Maâtouk et al. 1996 on low plasticity silts, *Bilotta & Foresta* on undisturbed pyroclastic soils). The changes in friction angle with suction are also considered by Toll (2000) on Kiunyu gravel (7% silt and 8% clay) that shows an increase of $\varphi^a$ (or $\varphi'$, angle in the $\tau$ vs $\sigma - u_a$ plane) with increased suction (Figure 26).

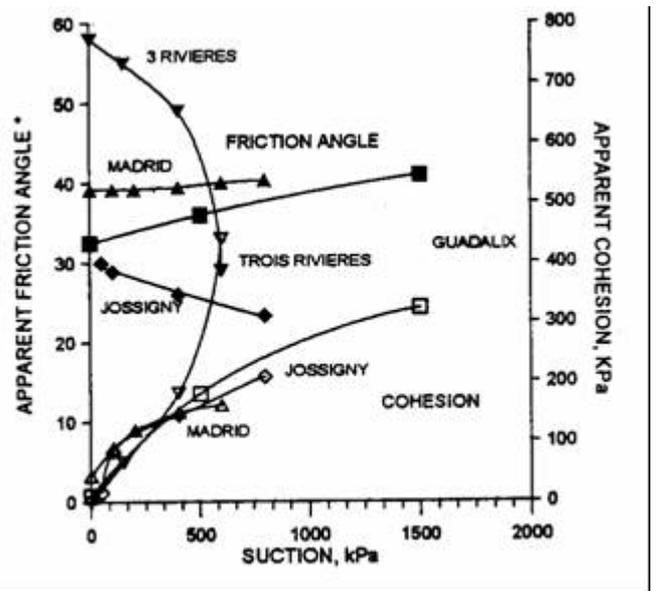

Figure 25 : changes in cohesion and friction angle with suction (Delage & Graham 1995)

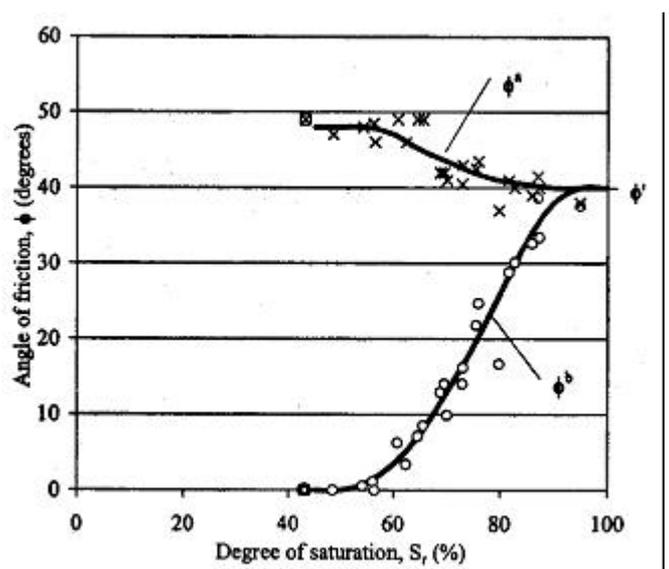

Figure 26 : changes in friction angle $\varphi^a$ with suction for Kiunyu gravel (Toll 2000)

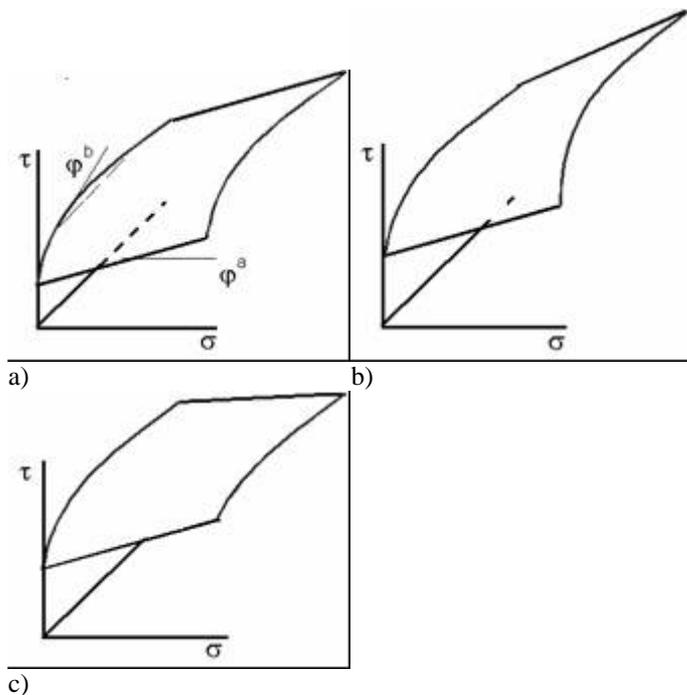

Figure 27 : Different shapes of surfaces representing the shear strength resistance of unsaturated soils : a) constant $\varphi'$ ; b) increasing $\varphi'$ with increased suction ; c) decreasing $\varphi'$ with increased suction.

Figure 27 presents the different possible surface in the ($\tau$ / $\sigma$ - $u_a$ / $u_a$ - $u_w$) plane, according to the various possible changes in $\varphi^a$ with suction.

In this Conference, *Vaunat, Romero, Marchi & Jommi* propose an approach to account for the effects of hysteresis on the shear strength properties of an unsaturated silt. Obviously, the effects of suction cycles and the resulting hysteresis on the shear strength properties of unsaturated soils appear to be significant and would deserve further investigation.

## 6   INNOVATIVE TRENDS

Among the various innovative approaches recently published, one notes the developments of suction controlled true triaxial testing of unsaturated soils (Matsuoka et al. 1998, Hoyos & Macari, Macari & Hoyos, 2001) and the use of image X ray tomography for the local determination of the degree of saturation in sand columns (Wong & Wibowo 2000).

Various contributions in the Conference deal with in-situ investigations in terms of determination of in-situ suction and/or moisture contents and water retention curves as a function of seasonal climate changes. These researches are of a considerable interest both in terms of geotechnical and geoenvironmental engineering to better understand and predict the responses of unsaturated soil structures sensitive to water changes and to ground-atmosphere exchanges (Blight 1997). Among other things, one can mention slope stability, the effects of drying seasons on buildings and related hazards, the behaviour of cover liners in surface waste disposals and the behaviour of embankments. In this Conference, contributions of *Paronuzzi, Melgarejo, Camapum, Vertamatti, Gerscovich, Li, Wang, Trichês, Bertolino, Phani-Kumar, Bilotta, Bastos* deal with these aspects, with some emphasis put on tropical soils. Other innovative contributions concern studies on the development of an air pressure calibration chamber (*Miller, Muraleetharan, Tan & Lauder*), erodability (*Bastos, Gehling, Bica & Milititsky*), cracking (*Avila, Ledesma & Lloret*), curling (*Nahlawi & Kodikara*), environmental concerns including chemistry effects (*Mata, Romero & Ledesma ; Li, Smith, Fityus & Sheng ; Shibata & Tukahara ; Proust, Jullien & Cui*) and cyclic behaviour (*Fleureau, Hadiwardoyo, Dufour-Laridan, Langlois & Gomez-Correia)*. TDR measurements of soil moisture in unsaturated soils seem to be more commonly used (*Mahler, Mendez, Souza & Fernandes ; Trichês & Pedrozo*).

## 7   CONCLUSION

In this general report, experimental systems and procedures of testing the mechanical behaviour of unsaturated soils have been presented. After describing the techniques of controlling suction, the water retention curves of unsaturated soils have been related to various parameters and physical properties.

In the last years, significant contributions concerning the water retention properties of compacted and natural (more often tropical soils) have been provided, and this topic is largely commented in this Conference. Relationship between water retention properties and hydraulic conductivity have also been more deeply investigated, with various assessments of the relevance of the numerous mathematical expressions of both parameters. Note that volume changes during suction cycles and the related hysteresis effects appear to be less documented and will probably deserve further investigation in the future, both on an experimental and conceptual point of view.

Different suction controlled oedometers and triaxial apparatuses have also been described and discussed in this report, together with the various existing system of monitoring volume changes. Some typical trends that illustrate the mechanical behaviour of unsaturated soils, in an elasto-plastic framework, have also been presented. In parallel, experimental techniques have been improved so as to allow for the precise monitoring of water exchanges in unsaturated soils submitted to changes in stress. Based on this, further investigation on the constitutive modelling of these water changes are presently being developed.

Recent contributions and this Conference have shown that many high quality experimental devices providing complete sets of data on the hydro-mechanical behaviour of unsaturated soils are now available. The data obtained can be used to help in geotechnical design that involve unsaturated soils, and to anticipate, or follow theoretical developments that can also help in geotechnical design. Presently, due to an increased interest in geoenvironmental engineering, the investigation of hydro-mechanical coupled effects is extended towards temperature and chemistry effects.